# Low cost and high performance light trapping structure for thin-film solar cells

DongLin Wang, Huijuan Cui & Gang Su*

Theoretical Condensed Matter Physics and Computational Materials Physics Laboratory, School of Physics, University of Chinese Academy of Sciences, P. O. Box 4588, Beijing 100049, China. Correspondence and requests for materials should be addressed to G.S. (email: gsu@ucas.ac.cn).

Nano-scaled dielectric and metallic structures are popular light tapping structures in thin-film solar cells. However, a large parasitic absorption in those structures is unavoidable. Most schemes based on such structures also involve the textured active layers that may bring undesirable degradation of the material quality. Here we propose a novel and cheap light trapping structure based on the prism structured $SiO_2$ for thin-film solar cells, and a flat active layer is introduced purposefully. Such a light trapping structure is imposed by the geometrical shape optimization to gain the best optical benefit. By examining our scheme, it is disclosed that the conversion efficiency of the flat a-Si:H thin-film solar cell can be promoted to exceed the currently certified highest value. As the cost of $SiO_2$-based light trapping structure is much cheaper and easier to fabricate than other materials, this proposal would have essential impact and wide applications in thin-film solar cells.

**Introduction**

Solar cells with high conversion efficiency and low cost are developmental goals of numerous studies. To realize such purposes, thin-film solar cells are promising and have tremendous advantages[1-3], in which a thinner active layer can reduce material usage and enhance the production rates, and can also minish the carrier collection length as well as the bulk recombination[4-6]. However, the poor light absorption in the thin active layer is a crucial bottleneck that we have to face and solve[7]. Multitudes of light trapping structures are therefore proposed to improve the light absorption in thin-film solar cells[8-14]. The most popular thin-film hydrogenated amorphous silicon (a-Si:H) solar cells have random pyramidal texture employed solar modules, and exhibit remarkable light trapping capabilities achieving the certified world-record conversion efficiency (10.2±0.3%).[1,15] In such cells, the periodically nanodome or nanocone plasmonic back reflectors[16-19], photonic crystals[20-23], whispering gallery resonator in spherical nanoshells[23] or nanospheres[24], and optical resonator in dielectric nanostructures[25-28] also reveal the outstanding light trapping. Nonetheless, in these schemes the involved nanostructured active layers may degrade the quality of materials due to the formation of localized defects around the nanostructures[16,18,29,30] that may lower the performance of the cell. To avoid the unexpected degradation, the development of the high performance light trapping structures is therefore urgent and significant for thin film solar cells.

Generally, a flat structure is beneficial to maintain the high quality of active material, but unfortunately, it is unsuitable for light trapping. Of the best is to seek a novel and cheap light trapping structure that is independent of the flat active layer but can provide a good optical benefit. Recently, it is reported that the periodically arranged plasmonic surface scatterers[31-34] on a flat a-Si:H solar cell can achieve 10% enhancement of photocurrent[31], but a large parasitic absorption in such metallic nanostructure can waste most of the trapped light. The high refractive indexed dielectric nanostructures are also proposed to trap light into the flat active layer by exciting optical resonances[26,27,35], and it was shown that although the geometrical shape optimization of the high-indexed GaP scatterers (refractive index n>3) on the top surface of a flat a-Si:H solar cell can improve the conversion efficiency by 22%, the large parasitic absorption of the GaP scatterer in the short wavelength region is still hard to avoid[27]. On the other hand, it was observed that a common but cheap material $SiO_2$ shows a prominent capacity for light trapping by whispering gallery modes in nanosphere arrays[24,36], which is free from the parasitic absorption and can achieve 11% enhancement of conversion efficiency for a flat a-Si:H solar cell[35]. It is not yet clear whether this material still has rooms for further improving the light trapping in thin-film solar cells.

Compared with other light trapping materials, $SiO_2$ is indeed much abundant and cheaper on earth, which has, however, a weaker capacity of light matter interaction than metal and high refractive indexed materials, and consequently, has limited applications in light trapping for thin film solar cells. Here, we demonstrate that, by properly optimizing its geometrical shape, this common material can have an excellent light trapping ability for thin film solar cells. As a practice, we employ a flat a-Si:H thin film solar cell as an example, and arrange the shaped $SiO_2$ on the front surface of the solar cell to trap light into the active layer, which gives unexpectedly that the conversion efficiency of the flat a-Si:H thin-film solar cell can be promoted to exceed the currently certified highest value. To design the architecture of the $SiO_2$ structure to achieve the maximum light trapping, an effective geometrical shape optimization algorithm combined into a full field optical and electrical simulation method[11,14,30,36-38] is employed.

**Results**

**Light trapping of prism structured $SiO_2$**. In a flat a-Si:H solar cell, the active layer less than 300nm is favorable for reducing the light-induced degradation[39-41]. In the short wavelength region ($\lambda$<550nm), such a thin layer can still absorb most of light in a single path due to the superior ability of light absorption of a-Si:H. The effective light trapping in this region is to eliminate the light reflection at the top surface of the solar cell. Conversely, in the long wavelength region ($\lambda$>550nm), the single light path is insufficient to absorb most of light. Thus, adjusting the direction of the incident light to enhance the light path in the active layer must be a feasible way. In general, there are four ways, the plasmonic resonator, the optical resonator, the whispering gallery resonator and the grating diffraction, to implement the light trapping in thin-film solar cells. The prism structured $SiO_2$ has both capacities of reducing the reflection and increasing the light path, and it is possible to design such a prism structure to achieve a dramatic light trapping.

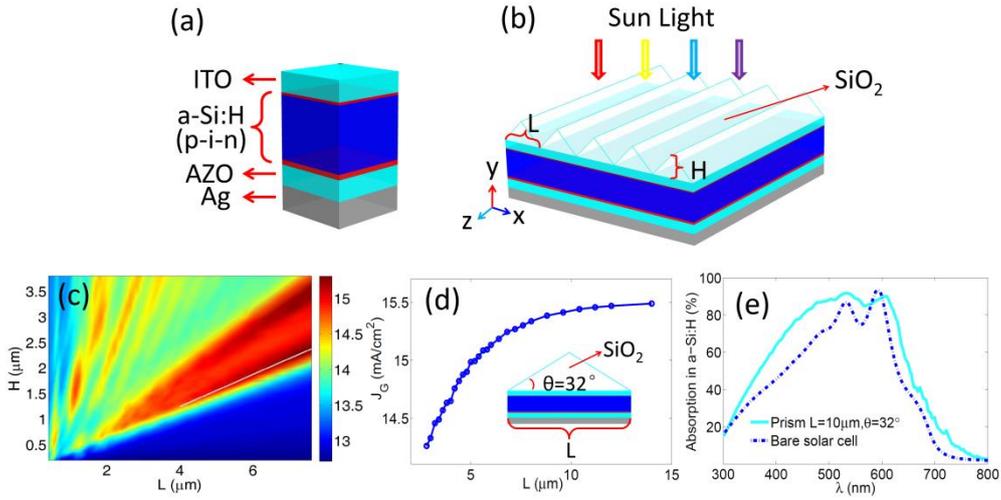

**Figure 1 | Illustration of the proposed structure and optical properties of bare and prism employed flat a-Si:H thin- film solar cells.** (a) The structures of the bare solar cell and (b) the solar cell with prism shaped $SiO_2$. (c) The density of photo-generated current ($J_G$) for the solar cell with different section sizes (H and L) of the prism structures. The inserted white line represents the prism with base angle θ=32°. (d) The $J_G$ of the solar cell with the prism angle θ=32° against the bottom width L of the prism. (e) Calculated light absorption efficiency in the active layer of the solar cell with L=10μm and θ=32°. The light absorption efficiency for the bare solar cell is taken as a reference.

In this work, a flat a-Si:H solar cell is employed to sustain the best material quality of the active layer. Figure 1a presents a flat solar cell[19] that contains 80nm thick indium doped tin oxide (ITO), 10nm thick *p* doped layer, 250nm intrinsic layer, 20nm *n* doped layer, 80nm thick aluminum doped zinc oxide (AZO) layer and 100nm thick Ag back reflector. We arrange the prism structured $SiO_2$ scatterers periodically on the top surface of such an a-Si:H solar cell (Figure 1b ). The cross-section of the prism is of an equilateral triangular geometry with height H and bottom width L. In this scheme, the prism based light trapping structure is distinct from the flat solar cell for only providing good optical benefit.

First, let us investigate how much optical benefit can be gained for the solar cell by employing such a prism structure. The optical property of the solar cell is assessed by the density of photo-generated current given by[42,43]

$$J_G = q \int \frac{A(\lambda) P_{am1.5}(\lambda) \lambda}{hc} d\lambda, \qquad (1)$$

where q is the charge of an electron, c is the speed of light, h is the Planck constant, A(λ) is the spectral absorption efficiency of the active layer, and $P_{am1.5}(\lambda)$ is the spectral photon flux density in solar spectrum (AM 1.5). For the bare a-Si:H solar cell, $J_G$ can reach 12.5 mA/cm². For the solar cells with different-sized prism based light trapping structures, $J_G$ is shown in Figure 2c for different H and L, which indicates that the enhancement of $J_G$ can be achieved for all prism employed solar cells. When the section size of the prism is small (H and L less than 1μm), the finite enhancement of $J_G$ comes principally from the antireflection effect that is caused by the prism generated graded-index coating. As the section size of the prism increases, the ray optics model is conventionally used to describe the mechanism of the light trapping in the prism structure. Hence, attuning the reflection and transmission at the interface of the prism is the main strategy to trap light into the active layer. Our calculation shows that the prisms with base angle θ from 29° to 46° are more suitable for the above purpose (Figure 1c). In particular, θ=32° of the prism is found to be the best choice to implement the light trapping in the flat solar cell.

A quantitative investigation of the light trapping capacity for the prism with θ=32° is presented in Figure 1d. The limit of $J_G$ for the solar cell with such a prism can reach 15.5 mA/cm$^2$ when the section size of the prism is large enough (L>15μm). To reduce the computational cost, the prism with L=10μm and θ=32° that can promote $J_G$ (15.43 mA/cm$^2$) of the solar cell close to the above limit is chosen as a fundamental light trapping structure for a further geometrical shape optimization. The light absorption efficiency in the active layer of the solar cell with such a structure is given in Figure 1e, which shows that an obvious enhancement of the light absorption can be obtained for almost all of wavelength from 300nm to 800nm. Of course, this fundamental prism structure can also provide much optical benefit for a flat solar cell by its good light trapping capacity.

**Geometrical shape optimization of the prism based light trapping structure**. The mechanism of the prism to enhance light absorption in the active layer is based on the control of the reflection and transmission at the interface between air and the prism. In the short wavelength region (λ<550nm), adjusting the side slope of the prism to reduce the reflection at the outside interface is an effective way for light trapping. Whereas, for the long wavelength (λ>550nm), the bottom Ag layer will reflect the unabsorbed light back to the interface. In this situation, it is a utility strategy to trap light by increasing the reflection at the inside interface to reflect the light back to the solar cell. As mentioned above, the fundamental prism structure with L=10μm and θ=32° can realize an excellent light trapping by balancing the light reflection at the interface of the prism side for both short and long wavelength. It is thus possible to realize a better balance for more light trapping by performing a geometrical shape optimization for this prism structure.

The geometry projection method (GPM)[44,45] is combined into a solar cell simulator[11,14,37] to optimize the shape of the prism structured SiO$_2$ for the best light trapping structure. GPM is an effective algorithm for two-dimensional shape optimization, which can generate smooth edge for the geometry that is convenient for practical fabrication. The solar cell simulator is based on a full field optical and electrical calculation that is carried out by solving Maxwell's equations, carriers transport equations and Poisson's equation (Details in Supplementary Note1).

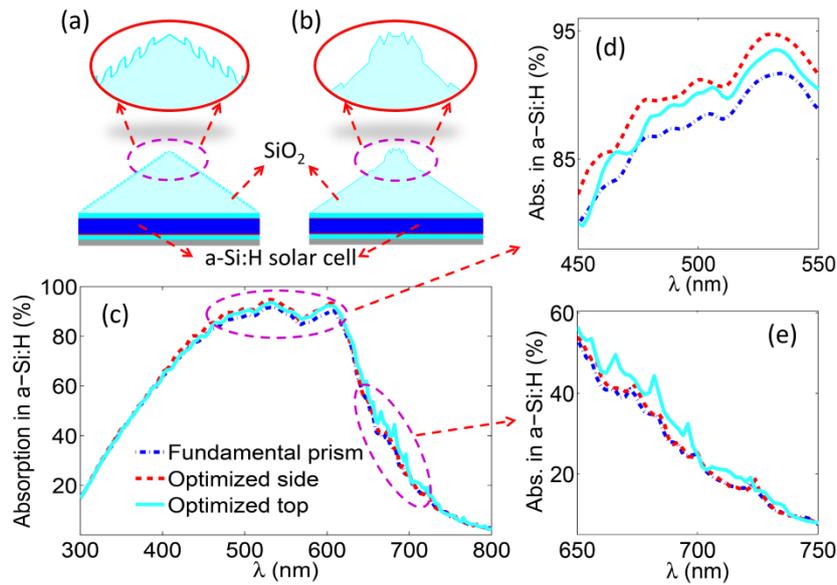

**Figure 2 | Structure of prism with optimized side and top as well as the corresponding light absorption efficiency in the active layer of the a-Si:H thin-film solar cell.** (a) The optimized side of the prism gives a zigzag structure. (b) The optimized top of the prism looks like a mountain shape. (c) Light absorption efficiency in the active layer of the solar cell with the prism of the optimized side and top. The fundamental prism is taken as a reference. Magnified light absorption efficiency in (c) for the wavelength region from 450nm to 550nm (d) and the wavelength region from 650nm to 750nm (e).

Generally, the light trapping of the prism is not uniform for short and long wavelength. Therefore, we have to employ different strategies to optimize the geometry of the prism. For the short wavelength, it is better to optimize the side shape of the prism to reduce the surface reflection. As shown in Figure 2a, a zigzag structure is introduced to generate a graded-index coating on the side of the fundamental prism for antireflection in the short wavelength region. Compared with the fundamental prism, the prism with optimized zigzag side can enhance the light absorption in the active layer from $\lambda=400$nm to $\lambda=600$nm (Figure 2c), and especially, the averaged enhancement can reach 3% from $\lambda=450$nm to $\lambda=550$nm (Figure 2d). For the light with long wavelength, we optimize the top geometry of the prism to adjust the incident direction to increase the light path by multiple reflections between the Ag layer and the inside surface of the prism. As shown in Figure 2b, the optimized top with a mountain shape of the prism has a better light trapping capacity than the fundamental prism from $\lambda=650$nm to $\lambda=750$nm (Figure 2c and 2e).

It is worth noting that the optimized side of the prism has no effect on the light trapping for the long wavelength (Figure 2e). In other words, the side with zigzag structure of the prism is insignificant to the reflection of its inside surface for the light with a long wavelength. The prism with an optimized top has an antireflective ability and can enhance the light absorption in the active layer for the light with a short wavelength (Figure 2d).

Therefore, the optical benefits provided by the optimized side and top of the prism can be superimposed to realize the best light trapping for the whole wavelength region (λ=300nm to λ=800nm). As shown in Figure 3a, the final optimized prism contains both zigzag side and mountain shaped top, which exhibits an excellent light trapping capacity for both short and long wavelengths. The solar cell with such an optimized prism can enhance obviously the light absorption in the active layer from λ=400nm to λ=700nm than the fundamental prism (Figure 3b). By gathering such enhancements, $J_G$ of the solar cell with the optimized prism can be improved to 16.36 mA/cm$^2$.

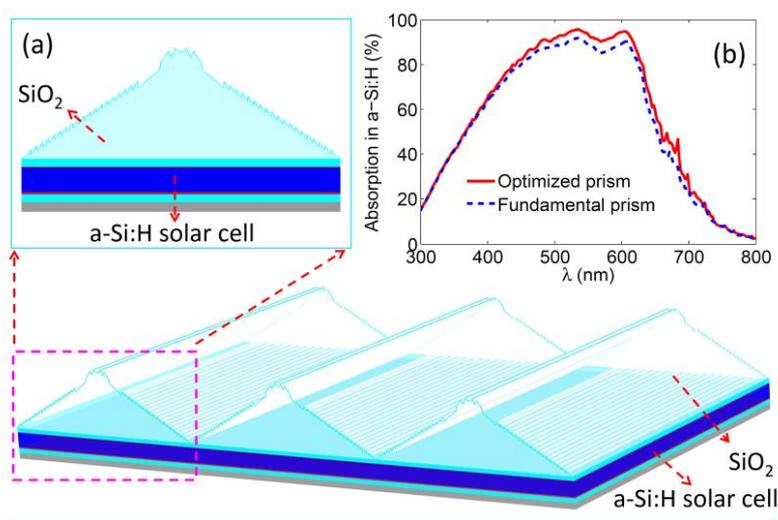

**Figure 3 | Structure of the optimized prism and the corresponding light absorption efficiency in the active layer of the a-Si:H thin-film solar cell.** (a) The optimized prism contains both zigzag side and mountain shaped top. The fine structure parameters can be found in Supplementary Note 2. The thickness of the a-Si:H solar cell is enlarged for eyes. (b) Light absorption efficiency in the active layer of the solar cell with the fundamental prism and the optimized prism.

**Electrical performance of the solar cell with proposed light trapping structures.** The fundamental and optimized prism structured SiO$_2$ can provide better optical benefit to trap more light into the a-Si:H thin film solar cell. Now, we want to show how much such optical benefit can be transformed into the electrical benefit of the solar cell. As shown in Figure 4, the current-voltage curves of the solar cell with or without light trapping structures are calculated by employing the solar cell simulator (Details in Supplementary Note 1). The calculated short circuit current density ($J_{sc}$=11.4 mA/cm$^2$) of the bare solar cell is consistent with the previous experiment[19]. This $J_{sc}$ of the solar cell has around 10% reduction than $J_G$ of the solar cell from the influence of carrier recombination. However, the remarkable light trapping capacity of the fundamental prism can still promote $J_{sc}$ of the solar cell to 14.01 (mA/cm$^2$), which leads eventually to that the conversion efficiency of the a-Si:H solar cell can be close to the certified highest record (10.2%). Moreover, the light trapping capacity of this fundamental prism can be further improved by employing the geometry shape optimization. Consequently, the solar cell with optimized prism has a higher $J_{sc}$ that can reach 14.88 (mA/cm$^2$). The conversion efficiency of the a-Si:H with such an optimized prism can be enhanced up to 10.74% that exceeds the certified highest efficiency, which is 32.6% larger than that for the bare solar cell. In addition, the so-designed light trapping structures have little influence on the open circuit voltage ($V_{oc}$) and the filling factor (FF) of the solar cells.

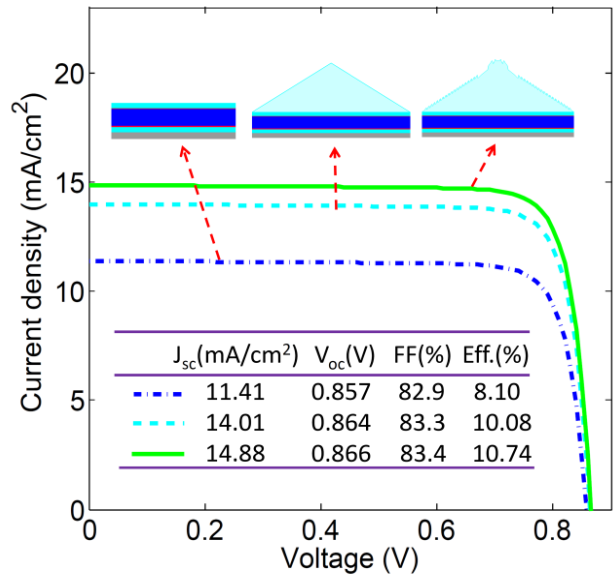

**Figure 4 | Current-voltage curve for the a-Si:H thin-film solar cells with and without prism based light trapping structures.** The short circuit current density ($J_{sc}$), the open circuit voltage ($V_{oc}$), the filling factor (FF) and the efficiency for the solar cell are listed for three cases.

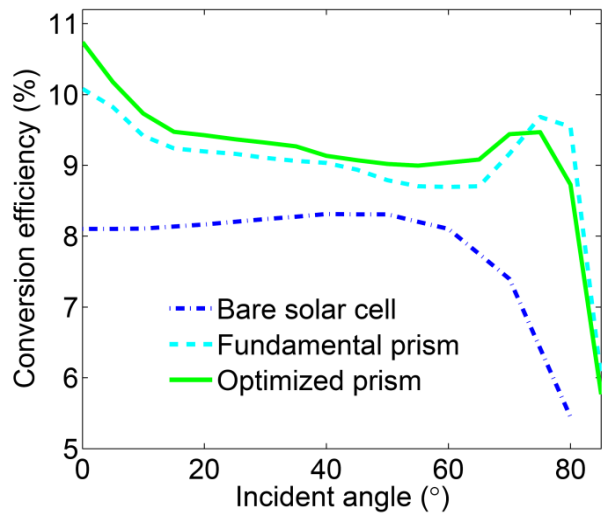

**Figure 5 | The conversion efficiency of the a-Si:H thin-film solar cells with and without prism based light trapping structures as function of the incident angle.**

Above results are all obtained with the normal incidence of light. We now examine the incident angle dependence for the solar cells with the designed light trapping structures. As shown in Figure 5, the enhancement of the conversion efficiency for the solar cells with the proposed structures of the prism can be sustained for all oblique angles. It can be seen that the serviceable angle of absorption light for the solar cell can be increased to about 80°. For the bare solar cell, the surface reflection will vanish for the transverse magnetic (TM) polarized incident light when the oblique angle of the light reaches Brewster's angle. This principle results ultimately in an obvious enhancement of the conversion efficiency for the bare solar cell when the incident angle around the Brewster's angle (~50°). For the prism employed solar cell, the incident light reaches the surface of the flat solar cell after it is refracted by the surface of the prism. It is understandable that the angle of the incident light is larger than that for the light actually arriving at the surface of the solar cell when the oblique angle is large. Therefore, the Brewster's effect for the solar cell with the prism based light trapping structure is moved to a larger incident angle (~80°). The practical Brewster's angle for the solar cell with the optimized prism structure is slightly smaller than that for the solar cell with the fundamental prism structure. This is because the refraction effect is weakened by the graded-index coating that is induced by the zigzag structure on the side of the optimized prism. Overall, the proposed light trapping structures can improve prominently the utilization of light for the oblique angle from 0° to 80°. In this angle region, the average conversion efficiency of 7.86% for the bare solar cell can be improved to 9.20% for the solar cell with the fundamental prism structure and 9.38% with the optimized prism structure.

**Discussion**

We proposed a novel strategy how to develop a high-performance and low-cost thin-film solar cells with designed light trapping structure implemented by cheap and common material $SiO_2$. Such a light trapping structure based on $SiO_2$ is so employed that can avoid parasitic absorption. It is structured by a prism shape arranging on the surface of a flat solar cell to provide good optical benefit. The geometrical shape optimization is applied to further improve the performance of the solar cell. This strategy is executed by combining the GMP and the full field optical and electrical simulations. By taking the a-Si:H thin-film solar cell as an example, we discover that the proposed prism based light trapping structure exhibits an excellent capacity to trap light into the cell. The calculated conversion efficiency of the solar cell with fundamental prism structure can be close to the reported highest value (10.2%). This impressive value can be further improved by optimizing the geometrical shape of the prism structure. The maximal efficiency of the a-Si:H thin-film solar cell with the optimized light trapping structure can be up to 10.74% that is 32.6% larger than that for the bare solar cell. Here we would like to stress that the material cost of the light trapping structure proposed in this work is much cheaper than other materials. In addition, it is quite possible that the proposed light trapping structure can be incorporated into other light trapping schemes (such as plasmonic back reflector) for further promoting the efficiency of the thin-film solar cells. The scheme proposed in this work is also applicable to the thin-film solar cell systems with the materials such as organometallic halide perovskite, cadmium telluride (CdTe), copper indium gallium selenide (CIGS) and so on.

**Methods**

The method is mainly composed of three modules: the geometrical shape optimization, the optical property calculation, and the electrical performance simulation[11,14,37,38]. The geometrical shape optimization of the light trapping structure is based on the GPM that can be used to simplify the traditional topological optimization problem. In the GPM, firstly, several control points are distributed on the optimization region; secondly, a three-dimensional (3D) surface can be fitted by the heights of those control points; and finally, the intersection curve between a level plane and the 3D surface can be used to define the geometrical shape of the designed structure. By employing this algorithm, we can transform the geometrical optimization problem into a simplified issue that seeks for the optimal heights of those control points. Then, the simplified optimization issue can be readily handled by Nelder-Mead algorithm[46]. Here, the GPM is combined into the finite element method (FEM) software package to construct the model of the solar cell[37,38]. After that, Maxwell's equations are solved by the FEM to realize the calculation of optical properties of the solar cell. The light field distribution in the active layer is extracted to describe the generation rate of the carriers. At last, the electrical performance simulation of the solar cell can be implemented by using the FEM to solve the transport equations of carriers and Poisson's equation. The calculational details can be found in Supplementary Note 1.

## Acknowledgments

The authors are benefitted from useful discussions with Q.B. Yan, Z.G. Zhu and Q.R. Zheng. This work is supported in part by the MOST of China (Grant No. 2012CB932900 and No. 2013CB933401), the Strategic Priority Research Program of the Chinese Academy of Sciences (Grant No. XDB07010100), and the China Postdoctoral Science Foundation (2014M550805).

## Author contributions

D.W. and G.S. conceived the project. D.W. and H.C. designed and executed the simulations. All authors prepared and contributed to the editing of the manuscript.

## Additional information

Supplementary Information accompanies this paper at http://www.nature.com/

**Competing financial interests:** The authors declare no competing financial interests.

## Figure Legends

**Figure 1 | Illustration of the proposed structure and optical properties of bare and prism employed flat a-Si:H thin- film solar cells.** (a) The structures of the bare solar cell and (b) the solar cell with prism shaped $SiO_2$. (c) The density of photo-generated current ($J_G$) for the solar cell with different section sizes (H and L) of the prism structures. The inserted white line represents the prism with base angle θ=32°. (d) The $J_G$ of the solar cell with the prism angle θ=32° against the bottom width L of the prism. (e) Calculated light absorption efficiency in the active layer of the solar cell with L=10μm and θ=32°. The light absorption efficiency for the bare solar cell is taken as a reference.

**Figure 2 | Structure of prism with optimized side and top as well as the corresponding light absorption efficiency in the active layer of the a-Si:H thin-film solar cell.** (a) The optimized side of the prism gives a zigzag structure. (b) The optimized top of the prism looks like a mountain shape. (c) Light absorption efficiency in the active layer of the solar cell with the prism of the optimized side and top. The fundamental prism is taken as a

reference. Magnified light absorption efficiency in (c) for the wavelength region from 450nm to 550nm (d) and the wavelength region from 650nm to 750nm (e).

**Figure 3 | Structure of the optimized prism and the corresponding light absorption efficiency in the active layer of the a-Si:H thin-film solar cell.** (a) The optimized prism contains both zigzag side and mountain shaped top. The fine structure parameters can be found in Supplementary Note 2. The thickness of the a-Si:H solar cell is enlarged for eyes. (b) Light absorption efficiency in the active layer of the solar cell with the fundamental prism and the optimized prism.

**Figure 4 | Current-voltage curve for the a-Si:H thin-film solar cells with and without prism based light trapping structures.** The short circuit current density ($J_{sc}$), the open circuit voltage ($V_{oc}$), the filling factor (FF) and the efficiency for the solar cell are listed for three cases.

**Figure 5 | The conversion efficiency of the a-Si:H thin-film solar cells with and without prism based light trapping structures as function of the incident angle.**

# Supplementary Information

# Low cost and high performance light trapping structure for thin-film solar cells

DongLin Wang, Huijuan Cui & Gang Su*

Theoretical Condensed Matter Physics and Computational Materials Physics Laboratory, School of Physics, University of Chinese Academy of Sciences, P. O. Box 4588, Beijing 100049, China.
*Correspondence and requests for materials should be addressed to G.S. (email: gsu@ucas.ac.cn).

**Supplementary Figure**

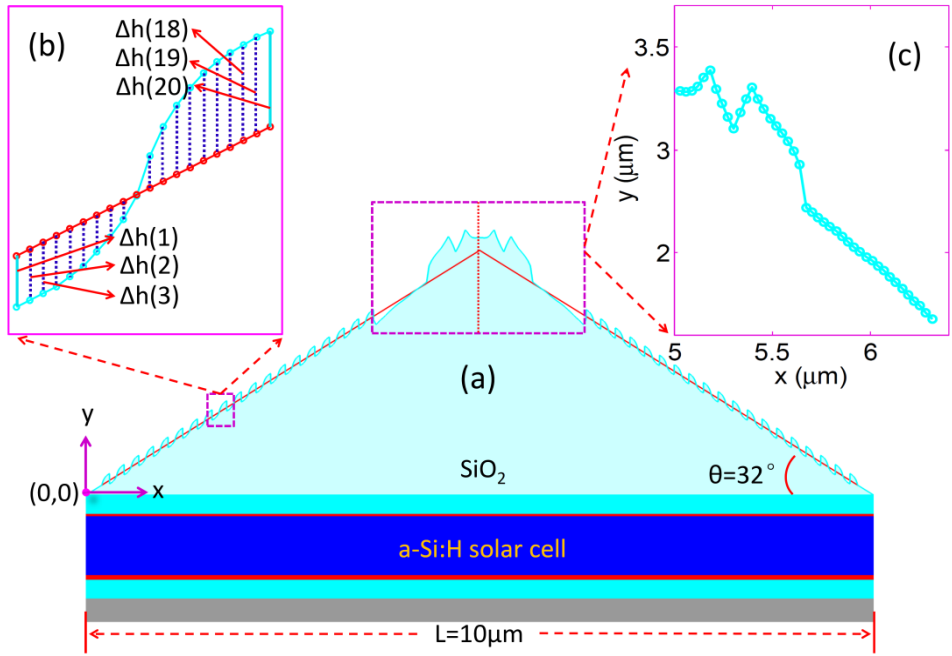

**Figure S1. Fine structure of the optimized prism structure.** (a) Side view of the structure for the fundamental prism and the optimized prism. The optimized prism structure contains zigzag sides and a mountain shaped top. (b) The fine structure of the zigzag structure. (c) The coordinates of the top structure of the optimized prism.

**Supplementary Table**

**Table S1:** Height difference between the edge of the fundamental prism and the optimized prism with zigzag sides.

| $\Delta h$ | \multicolumn{10}{c}{The unit of data is 10nm} |
|---|---|---|---|---|---|---|---|---|---|---|
| $\Delta h(1) \sim \Delta h(10)$ | -4.6 | -4.6 | -4.6 | -4.6 | -4.3 | -4.0 | -3.3 | -2.6 | -1.6 | -0.1 |
| $\Delta h(11) \sim \Delta h(20)$ | 2.9 | 4.9 | 6.2 | 7.2 | 7.9 | 8.3 | 8.5 | 8.6 | 8.7 | 8.6 |

## Supplementary Note 1

**Computational details**

In this work, the geometry projection method (GPM)[44,45] is employed to implement the geometrical shape optimization of the light trapping structure. To improve the efficiency and accuracy of calculations, the two-step optimization method is used. In the first step, the standard GPM is adopted to seek for a coarse result of the optimal light trapping structure. The edge of this coarse structure is undulated which relies on the square mesh structure in the algorithm. In the second step, the above coarse result is used as the initial structure in the improved GPM to search for the refined structure for the best light trapping. In the improved GPM, the intersection curve between the 3D surface and the level plane is directly extracted to define the shape of the optimal structure. The intersection curve is approximated by a polygon with 20~50 sides for accurately describing the optimal structure. The data of those sides are transformed into the finite element method (FEM) script to realize the data connection between the GPM and FEM software package. The mirror symmetry is imposed to the prism structure to reduce the computational costs

The optical property of the solar cell is simulated by solving Maxwell's equations in FEM software package[37,38]. All of optical calculations are implemented under a normal incidence unless specified. Because the prism based light trapping structure is asymmetrical in $x$ and $z$ directions, both the transverse electric (TE) and the transverse magnetic (TM) polarized incident light are considered. The final calculations give the averaged results for TE and TM modes. The complex optical constants for AZO, ITO, AZO, Ag, a-Si:H and $SiO_2$ are taken from previous experimental works[47-50]. By performing the optical simulation, we can obtain the optical absorption and light field distribution in each layer of the solar cell.

The simulation method to implement the electrical performance is based on a previous work[37,38]. It is carried out by solving the transport equations of carriers and Poisson's equation by using the FEM software package. The only difference here is that more accurate generation profile is used, which is expressed as[14]:

$$G(\vec{r},\lambda) = \frac{\varepsilon'' |E(\vec{r},\lambda)|^2}{2\hbar} \quad (S1)$$

where $E(\vec{r},\lambda)$ is the distribution of the electric field intensity at each single wavelength in the active layer, $\varepsilon''$ is the imaginary part of permittivity of a-Si:H. This generation profile should be weighted by AM1.5G sun spectrum[51] to simulate one sun illumination. The simulation details of the electrical performance can be found in previous reports[37,38], and the electrical parameters of a-Si:H are taken from previous studies[38,52]. Besides, 5 $\Omega cm^2$ series resistance and 5 $k\Omega cm^2$ shunt resistance are applied to the a-Si:H solar cell[30].

## Supplementary Note 2

**The parameters of the fine structure of the optimized prism**

The geometrical shape optimization is applied to the fundamental prism structure that has an isosceles triangle section with bottom width L=10μm and base angle θ=32° (Figure S1).

Based on this fundamental prism, an optimal prism structure that has a better light trapping capacity is obtained by the optimization calculation. The optimized prism structure contains both zigzag sides and a mountain shaped top. As shown in Figure S1a, the zigzag structure is periodically arranged on the side of the prism. There are 19 zigzag periods in total on one side of the prism, and the width of one zigzag period is 185nm. For an accurate description of this structure, we divide equally one unit structure into 19 sections along the $x$ direction to generate 20 nodes on the edge of the zigzag structure. As indicated in Figure S1b, the height difference between the side of the fundamental prism and 20 nodes on the edge of the zigzag structure is denoted as Δh(1)~Δh(20). Those Δh are listed in Table S1. By using the same method, a half of the mountain shaped top of the optimized prism is equally divided into 43 sections along the $x$ direction. By setting the left corner of the prism as the coordinate origin (0,0), the half of the mountain structured top can be accurately described by the coordinates of 44 nodes on its edge (Figure S1c).

**Supplementary References**